\documentstyle{laa}     
\newcommand{\gcc}{$g~cm^{-3}\ $}

\newcommand{\msun}{$M_{\odot}\ $}

\begin{document}
\input psbox.tex
 
   \thesaurus{02.05.2; 02.10.8; 08.10.1; 08.14.1; 08.16.6}
   \title{Equilibrium Sequences of Rotating Neutron Stars for New Realistic 
     Equations of State}
 
   \author{ B. Datta \inst{1} \thanks{Also at Raman Research Institute,
            Bangalore 560 080, INDIA}, A. V. Thampan \inst{1}and I. Bombaci
          \inst{2}
          }

   \institute{$^1$ Indian Institute of Astrophysics, Bangalore 560 034,INDIA.\\
              $^2$ Dipartimento di Fisica, Universit\'{a} di Pisa, and INFN Sezione 
              di Pisa, Piazza Torricelli 2, I-56100 Pisa, ITALY}

   \date{}
 
   \maketitle
 
   \begin{abstract}
For four newly suggested realistic equations of state of neutron star matter,
we construct equilibrium sequences of rapidly rotating neutron stars in
general relativity.  The sequences are the normal and supramassive evolutionary
sequences of constant rest mass.  We find that for these equations  of state
the maximum (gravitational) mass rotating models occur (in central density
and rotation rate $\Omega$) before the maximum--$\Omega$ models.  We calculate
equilibrium sequences for a constant value of $\Omega$ corresponding to the 
most rapidly rotating pulsar PSR 1937+21.  Also calculated is the radius of
the marginally stable orbit and its dependence on $\Omega$, relevant for 
modeling of kilo--Hertz quasi--periodic oscillations in X--ray binaries.
\end{abstract}


\section{Introduction}

Equilibrium sequences of rapidly rotating neutron stars are important
in modelling a variety of phenomena of astrophysical interest, such as 
millisecond pulsars, low--mass X--ray binaries (LMXBs) and Quasi Periodic 
Oscillators (QPOs).  Models of rapidly rotating neutron stars in general 
relativity must be constructed numerically.  Early work on this have been
based on incompressible fluids and polytropic models (Bonazzola \& Schneider
1974; Butterworth 1976).  In 1986 Friedman et al. (1986) reported calculations 
of rapidly rotating neutron stars in general relativity using a set of 
realistic equations of state (EOS) for neutron star matter. A similar work 
based on a formalism due to Komatsu et al. (1989) (KEH formalism) was done by
Cook et al. (1994) for purpose of studying quasi--stationary evolution of 
isolated neutron stars.  An alternative approach based on spectral methods was 
developed by Bonazzola et al. (1993).  Extensive calculations using the 
spectral method for a broad set of realistic EOS of neutron star matter were
presented in Salgado et al. (1994a; 1994b).

A key input in determining the structure of neutron stars is the EOS of high
density matter. The work of Friedman et al. (1986) and that of Cook et al. 
(1994) make abundantly clear that the EOS also plays a significant role in 
deciding the various equilibrium sequences of rotating neutron stars.  For 
example, the Keplerian frequency of a test particle in orbit around a neutron 
star ranges from 55\% of its spherical value for models based on the softest 
EOS to 75\% of the spherical value for models with the stiffest EOS.  The 
spreads in rotation--induced changes in the values of masses and radii from 
static neutron star cases also display considerable EOS dependence.  These 
quantities (especially the Keplerian frequency of a particle in orbit around 
the rotating neutron star and the radius of the innermost stable circular orbit)
are important for deciding the boundary layer structure and hence the emission 
characteristic of LMXB and QPOs.

Although dense matter has been a subject of study for nearly three decades now, 
there is no general agreement still on its exact composition, and on its EOS, 
especially for densities in excess  several times nuclear matter density. 
The bulk of a neutron star ({\it core}) is made up of an electrically 
neutral quantum fluid composed of neutrons, protons, electrons and muons in 
equilibrium with respect to the weak interactions (beta--stable nuclear matter).    
However, at ultra--high densities, a variety of new and exotic hadronic 
degrees of freedom may become important (like hyperons, a $K^-$ condensate or a 
deconfined phase of quark matter). The possible appearance of such an 
exotic core has enormous  consequences for the transport properties of neutron 
stars and also for the formation of black holes (Brown \& Bethe 1994; 
Bombaci 1996).    
The consequences of the existence of an exotic core (such as quark matter or 
kaon condensation) on the properties of rapidly rotating neutron stars will be 
reported in a forthcoming paper (Datta et al. 1997). In the present work, we 
have considered a conventional picture assuming the neutron star core to be 
composed of only beta--stable nuclear matter.  Even in this picture, the 
determination of the EOS of asymmetric nuclear matter to describe the core 
of the neutron star, remains a formidable theoretical problem.  
In fact, one has to extrapolate the EOS to extreme conditions of 
high density and high neutron--proton asymmetry, {\it i.e.} in a regime 
where the EOS is poorly constrained by nuclear data and experiments. 
Astrophysical observational data, such as based on the binary pulsar 
PSR 1913+16, which give $1.4$~\msun as the mass of the neutron star in the
binary (Taylor \& Weisberg 1989) and analysis of Vela pulsar postglitch 
timing data can be used to broadly rule out very soft EOS (Datta \& Alpar 1993).
Recently, some new EOS of asymmetric nuclear matter have been calculated 
and applied to the study of non--rotating neutron stars (Baldo et al. 1997, 
Bombaci 1995).  These EOS are based on  (i) a microscopic 
Brueckner--Bethe--Goldstone many--body approach and 
(ii) a phenomenological model based on effective nuclear forces.  
These satisfy the basic requirements of reproducing the empirical saturation 
point for  symmetric nuclear matter, the symmetry energy and the 
incompressibility parameter at the saturation density (see Table 1). 
These models have the desirable physical feature that the velocity of sound in 
the medium does not violate the causality condition. Therefore, these can be 
taken to be {\it realistic} EOS, and so it would be of interest
to see the equilibrium rotating sequences that would be possible with these 
EOS. In this paper we report calculations of equilibrium sequences of 
rapidly rotating neutron stars in general relativity for these new realistic 
EOS models.  The various equilibrium sequences that we construct are
 normal and supramassive evolutionary sequences of constant rest 
mass. In addition, we build equilibrium sequences for a constant value of
rotation rate corresponding to a period of $P= 1.558$~ms of 
the millisecond pulsar PSR 1937+21 (Backer et al. 1982), the most rapidly 
rotating pulsar known.

\section{ Rapidly and Rigidly Rotating Relativistic Stars}

The space--time around a rotating neutron star can be described in 
quasi--isotropic coordinates, as a generalization of Bardeen's metric
(Bardeen 1970):

\begin{eqnarray}
ds^2 & = & g_{\mu \nu}dx^{\mu}dx^{\nu} (\mu, \nu = 0,1,2,3) \nonumber \\
     & = & -e^{\gamma+\rho} dt^2 + e^{2\alpha} (r^2d\theta^2 + dr^2) +
         e^{\gamma-\rho} r^2 sin^2\theta \nonumber \\
     &   & ~~~~~~~~~~~~~  (d\phi - \omega dt)^2 
\end{eqnarray}

\noindent where $g_{\mu \nu}$ is the metric tensor.  The metric 
potentials $\gamma$, $\rho$, $\alpha$, and the
angular velocity of the stellar fluid relative to the local inertial frame
($\omega$) are all functions of the quasi--isotropic radial coordinate ($r$) 
and the polar angle ($\theta$).  We use here geometric units: $c=1=G$.
We assume a perfect fluid description, for which the energy momentum tensor 
is given by:

\begin{eqnarray}
T^{\mu \nu} & = & (\epsilon + P)u^{\mu}u^{\nu} + P g^{\mu \nu}
\end{eqnarray}

\noindent where $\epsilon$ is the total energy density, $P$ the pressure 
and $u^{\mu}$ the unit time--like four velocity vector that satisfies

\begin{eqnarray}
u^{\mu}u_{\mu}& = & -1
\end{eqnarray}

The proper velocity $v$ of the matter, relative to the local Zero Angular 
Momentum Observer (ZAMO), is given in terms of the the angular velocity 
$\Omega \equiv u^3/u^0$ of the fluid element (measured by a distant 
observer in an asymptotically flat space--time), by the following equation
(see Bardeen 1970):

\begin{eqnarray}
v & = & (\Omega-\omega) r sin\theta e^{-\rho}
\end{eqnarray}

\noindent The four velocity ($u^{\mu}$) of the matter can be written as

\begin{eqnarray}
u^{\mu} = \frac{e^{-(\gamma+\rho)/2}}{(1-v^2)^{1/2}} (1,0,0,\Omega)
\end{eqnarray}

Substitution of the above into Einstein field equations projected on to the 
frame of reference of a ZAMO yield three elliptic equations for the metric 
potentials $\rho$, $\gamma$ and $\omega$  and two linear ordinary differential 
equations for the metric potential $\alpha$ (Komatsu et al. 1989; 
Butterworth \& Ipser 1976; Bardeen \& Wagoner 1971).  In 
the  KEH formalism (Komatsu et al. 1989), the elliptic 
differential equations  are converted to integral equations for the metric 
potentials using Green's function approach.

From the relativistic equations of motion, the equation of hydrostatic 
equilibrium for a barytropic fluid may be obtained as:

\begin{eqnarray}
h(P)-h_p & \equiv &  \int_{P_p}^{P}\frac{d P}
{(\epsilon+P)} = {\rm ln} u^t - {\rm ln} u^t_p - 
\int_{\Omega_c}^{\Omega} F(\Omega)d\Omega
\end{eqnarray}

\noindent where $h(P)$ is termed as the specific enthalpy.  $P_p$,
$u^t$ are the rescaled values of pressure and  t-component of the four
velocity respectively and $h_p$ is the specific enthalpy at the pole;  
$F(\Omega) = u^tu_{\phi}$ is the integrability condition
imposed on the equation of hydrostatic equilibrium, and it can be physically
interpreted as the rotation law for the matter constituting the neutron star. 
An appropriately chosen value of $h_p$ 
defines the surface of the star.  Equation (6) shows that the hydrostatic 
equilibrium equation is integrable if $P(\epsilon)$ and 
$u^tu_{\phi}$ are specified.

As shown by Bardeen 1970 (see also Butterworth \& Ipser 1976), the quantity 
 $u^tu_{\phi}$ is a function of $\Omega$ only. Komatsu et al. (1989) have 
suggested the following specific form for $F(\Omega)$:

\begin{eqnarray}
F(\Omega) & = & A^2 (\Omega_c - \Omega)
\end{eqnarray}

\noindent where $A$ is a rotation constant such that when $A\rightarrow\infty$,
the configuration approaches rigid rotation (that is, 
$\Omega=\Omega_c$) so that  $F(\Omega)$ is finite.
Furthermore, when $A\rightarrow0$, the configuration should 
approach that of rotation with constant specific angular momentum.

On substituting Eqs. (5) and (7) into Eq. (8), we have the 
hydrostatic equilibrium equation as 

\begin{eqnarray}
h(P) - h_p & = & \frac{1}{2}\left[\gamma_p + \rho_p - \gamma - \rho \right .
- {\rm ln}(1-v^2) + \nonumber \\
           &   & ~~~~~~~~~~~~~~~~~\left. A^2(\Omega - \Omega_c)^2\right]
\end{eqnarray}

\noindent where $\gamma_p$ and $\rho_p$ are the values of the metric potentials
at the pole, and $\Omega=r_e \Omega$.

Therefore, the hydrostatic equilibrium equations at the centre and equator for 
a rigidly rotating neutron star become respectively

\begin{eqnarray}
& & h(P_c) - h_p - \frac{1}{2}\left[\gamma_p + \rho_p - \right .
\left . \gamma_c - \rho_c \right ] = 0 \\
& & (\gamma_p + \rho_p - \gamma_e - \rho_e )
 -  {\rm ln}[1-(\Omega_e-\omega_e)^2r_e^2e^{-2 \rho_e}] = 0
\end{eqnarray}

\noindent where the subscripts p, e and c on the variables stand respectively 
for the corresponding values at the pole, equator and center.

We solve (numerically) the integral equations for $\rho$, $\gamma$ and 
$\omega$, the ordinary 
differential equation (in $\theta$) for the metric potential $\alpha$,
together with Eqs. (8), (9) and (10), iteratively to obtain $\rho$, $\gamma$, 
$\alpha$, $\omega$, the equatorial coordinate
radius ($r_e$), angular velocity ($\Omega$), and the density ($\epsilon$)
and pressure ($P$) profiles.


\subsection { Innermost stable orbits}

Since the metric (1) is stationary and axisymmetric, the energy and angular 
momentum are constants of motion.  Therefore, for a particle in stable orbit
around the neutron star, the specific energy $E$ 
(in units of the rest energy $m_{0}c^2$, where $m_{0}$ is the rest mass of the 
particle) and the specific angular momentum $l$ (in units of $m_{0}c$) 
can be identified as $-p_{0}$ and $p_{3}$ respectively, where, $p_{\mu}$ 
($\mu = 0, 1, 2, 3$), stands for the four--momentum of the particle.  
From the condition $p_{\mu}p^{\mu} = -1$, we have the equations of motion  
of the particle (confined to the equatorial plane) in this gravitational field 
as 

\begin{eqnarray}
\dot{t} & = & \frac{dt}{d\tau} =  p^0  = e^{-(\gamma + \rho)} (E - 
\omega l) \\
\dot{\phi} & = & \frac{d\phi}{d\tau} =  p^3 = \Omega p^0 = 
e^{-(\gamma+\rho)} \omega (E - \omega l) + 
\frac{l}{r^2 e^{(\gamma-\rho)}} \\
\dot{r}^2 & \equiv & e^{2\alpha + \gamma + \rho}
 \left(\frac{dr}{d\tau}\right)^2 = 
E^{2} - V^2 .
\end{eqnarray}

\noindent Here, $\tau$ is the proper time and $V$ is the effective
potential given by

\begin{eqnarray}
V^2 & = & e^{\gamma + \rho}
\left[1 + \frac{l^2/r^2} {e^{\gamma - \rho}}\right] 
+ 2\omega El - \omega^{2}l^2.
\end{eqnarray}

\noindent The conditions for circular orbits, extremum of energy and minimum 
of energy are respectively: 

\begin{eqnarray}
E^2 & = & V^2 \\
V_{,r} & = & 0 \\
V_{,rr} & > & 0.
\end{eqnarray}

\noindent For marginally stable orbits,

\begin{eqnarray}
V_{,rr} & = & 0.
\end{eqnarray}

\noindent In our notation, a comma followed by one `$r$' represents a first order 
partial derivative with respect to $r$, etc..

From the expression for the effective potential and the conditions (15), (16) 
and (17), one obtains three equations in the three unknowns: namely, $r$,
$E$, and $l$.  In principle, if analytical expressions for 
$e^{\gamma + \rho}$, $e^{2\alpha}$, $e^{\gamma - \rho}$ and $\omega$ are known, 
it would be a straightforward exercise to solve these equations to obtain $r$, 
$E$, and $l$. In practice, however, this is not so, and the 
solutions for the metric coefficients $e^{\gamma + \rho}$, $e^{2\alpha}$, 
$e^{\gamma - \rho}$, and $\omega$ have to be  obtained as arrays of numbers 
for various values of 
$r$ and $\theta$ using a numerical method. Furthermore, the condition (18) 
will introduce second order derivatives of $\gamma$, $\rho$, and $\omega$, 
which means that care has to be exercised in ensuring the numerical accuracies 
of the quantities calculated.  For this purpose, it is convenient to express 
$E$ and $l$ in terms of the physical velocity $v$ using Eq. (4) 
(Bardeen 1972) as:

\begin{eqnarray}
E-\omega l & = & \frac{e^{(\gamma+\rho)/2}}
{\sqrt{1 - v^2}}\\
l & = & \frac{v r e^{(\gamma-\rho)/2}}
{\sqrt{1 - v^2}}.
\end{eqnarray}

\noindent Eqs. (19) and (20) can be recognized as the condition
 for circular orbits.  Conditions (16) and (18) yield respectively,

\begin{eqnarray}
v & = & \pm \left(\sqrt{e^{-2\rho}r^4\omega_{,r}^{2} + 2r(\gamma_{,r} + 
\rho_{,r}) + r^2(\gamma_{,r}^2-\rho_{,r}^2)} \pm \right. \nonumber \\
& & ~~~~~~~~~~~~~ \Big. e^{-\rho} r^{2} \omega_{,r} \Big)/
(2 + r(\gamma_{,r} - \rho_{,r}) \\
V_{,rr} & \equiv & 2\left[\frac{r}{4}(\rho^2_{,r}-\gamma^2_{,r}) - 
\frac{1}{2}
 e^{-2\rho}\omega_{,r}^{2}r^3 - \rho_{,r} + \frac{1}{r}\right]v^2 
\nonumber \\
& &  + [2 + r(\gamma_{,r} - \rho_{,r})]vv_{,r} 
 - e^{-\rho}\omega_{,r} r v 
\nonumber \\
& & + \frac{r}{2}(\gamma^2_{,r} - \rho^2_{,r}) - e^{-\rho} r^2 \omega_{,r} 
v_{,r} = 0
\end{eqnarray}

\noindent where we have made use of Eq. (21) and its derivative with respect
to r in order to eliminate the second order derivatives in Eq. (22). 
The zero of $V_{,rr}$ will give the innermost stable circular orbit radius
($r_{orb}$).  In Eq. (21), the positive sign refers to the co--rotating 
particles and the negative sign to the counter--rotating particles.  In this  
study we have considered only the co--rotation case.


\section {Numerical Procedure}

The numerical procedure followed by us is the  KEH  formalism. This is based 
on an earlier work
by Hachisu (1986) which has a self--consistency requirement that requires that 
the maximum (central) energy density $\epsilon_c$ and  the ratio of the polar 
to equatorial radial coordinates $r_p/r_e$ be fixed for each iterative cycle.
If $\rho^i$, $\gamma^i$ , $\alpha^i$ and $r^i_e$ are the values
of the corresponding parameters during the $i^{th}$ iterative cycle, then:

\newcounter{add}
\begin{list}%
{\arabic{add}. }{\usecounter{add}
 \setlength{\rightmargin}{\leftmargin}}
\item these values are first scaled (divided) by $(r^i_e)^2$ to obtain
      $\hat{\rho}^i$, $\hat{\gamma}^i$ and $\hat{\alpha}^i$ respectively.
\item a new  value of $r_e$ is calculated using Eq. (10) for 
      $\epsilon=\epsilon_c$ i.e. $v=0$ so that

      \begin{eqnarray}
      r^{i+1}_e & = & \frac{2[h(P(\epsilon_c)) - h_p]}
      {\hat{\gamma}^i_p + \hat{\rho}^i_p - \hat{\gamma}^i_c - \hat{\rho}^i_c}
      \end{eqnarray}

\item the value of $\Omega_c$ is computed from Eq. (11) as

      \begin{eqnarray}
      \Omega^{i+1}_c & = &  \hat{\omega}^i_e + e^{\rho^i_e}
      \left[1 - e^{(\gamma^i_p + \rho^i_p - \gamma^i_e - \rho^i_e)}\right]
      \end{eqnarray}

\item the values of the three scaled metric potentials $\hat{\rho}^i$, 
      $\hat{\gamma}^i$ and $\hat{\alpha}^i$ are rescaled (multiplied)  by
      $(r^{i+1}_e)^2$

\item using these values of $r^{i+1}_e$, $\Omega^{i+1}_c$, 
      $\rho^i$, $\gamma^i$, $\alpha^i$, $\hat{\omega}^i$, equation (9) is 
      solved to obtain the matter energy distribution namely 
      $\epsilon^{i+1}$, $P^{i+1}$, $v^{i+1}$ etc.

\item the integral equations for the metric potentials are solved to obtain 
      $\rho^{i+1}$, $\gamma^{i+1}$, $\hat{\omega}^{i+1}$ and $\alpha^{i+1}$.

\item steps (1) to (6) are repeated until $r_e$ converges to within
      a tolerance of $10^{-5}$.
\end{list}

\noindent Once $r_e$ converges, the metric potentials $\rho$, $\gamma$,
$\omega$ and $\alpha$ together with the density ($\epsilon$) and pressure 
($P$) profiles can be used to compute the structure parameters (see 
Cook et al. 1994).

\section {New Equation of State Models}

\subsection { Microscopic equation of state }

  In a microscopic approach the input is the two--body nucleon--nucleon (NN) 
interaction, described by so--called {\it realistic} interactions like the 
Argonne, Bonn, Nijmegen, Paris, Urbana potentials 
(see {\it e.g.} Machleidt 1989).  
The theoretical basis to construct these realistic NN potentials is the 
meson--exchange theory of nuclear forces. In this scheme, nucleons, nucleon 
resonances (e.g. $\Delta(1232)$), and mesons such as $\pi$, $\rho$ and 
$\omega$, are incorporated in a potential representation. The various 
parameters in the potential are then adjusted to reproduce the experimental 
data for the two--body problem (deuteron properties and NN scattering phase 
shifts). 
Then one has to solve the complicated many--body problem to get the EOS.  

 Recently, Baldo et al. (1997), hereafter BBB, have 
computed a new EOS of beta--stable nuclear matter, and with this EOS they 
have calculated the structure of non--rotating neutron stars. 
In their approach, the energy per nucleon of nuclear matter is obtained in the 
Brueckner--Hartree--Fock (BHF) approximation of the 
Brueckner--Bethe--Goldstone theory. The only input quantity for these 
calculations is the nuclear interaction.  In their calculation BBB used the 
Argonne v14 (Av14)) (Wiringa et al. 1984) or the Paris (Lacombe et al. 1980) 
two--body nuclear force, implemented in both cases by the Urbana three-body 
force (TBF) (Carlson et al. 1983; Schiavilla et al. 1986).  As is well known, 
the need for a TBF arises to obtain a correct saturation point of symmetric 
nuclear matter in a non--relativistic many--body approach. 
In the following we refer to the  EOS obtained in BBB  with the 
Av14+TBF and Paris+TBF, as the BBB1 and BBB2 equation of state respectively. 
The saturation properties for these two microscopic models are summarized 
in Table 1, and the calculated speed of sound is shown in Fig. 1.  The latter
always remains within the causality bound.

\subsection { Phenomenological equation of state }

 In this case the input is a density-dependent effective NN interaction. 
The most popular of this kind of interaction is the Skyrme interaction 
(Skyrme 1956).  In the present work we used a generalized Skyrme--like EOS 
developed in ref. (Bombaci 1995), and we refer to it as the BPAL EOS. 
 An important feature of the BPAL models is the possibility to have different 
forms for the density dependence of the potential part $E_{sym}^{pot}(n)$  
of the nuclear symmetry energy, modelling different results predicted by 
microscopic calculations (Wiringa et al. 1988, Bombaci \& Lombardo 1991).  
In particular, $E_{sym}^{pot}$ is proportional to the nucleon number 
density $n$ in the case of BPAL32, and to $\sqrt{n}$ in the case of BPAL21 EOS.  
The density dependence of the symmetry energy  plays a very important role in 
the physics of neutron stars. 
This function determines the proton fraction in beta--stable
nuclear matter, which, in turn, is crucial for an accelerated rate of cooling
of a neutron star through the so-called direct Urca process above a critical 
value of the proton fraction (Lattimer et al. 1991; Page \& Applegate 1992).

In Fig. 2 we plot the pressure--density relationship for EOS models
BBB1, BBB2, BPAL21 and BPAL32 (curves 1 -- 4) and also compare it with two 
other EOS models, one of which is a {\it very soft} EOS BPAL12 (Bombaci 1995) 
and the other a {\it very stiff} EOS (Sahu et al. 1993):

(1) Soft: {\it BPAL21} ({\it Bombaci 1995}): This EOS is characterized by 
$K_0 = 120$ MeV and $E_{sym}^{pot} \sim n$. The value 120 MeV for the 
 incompressibility is unrealistically small when compared with the value 
$220 \pm 30$ MeV extracted from nuclear phenomenology (Blaizot 1980; 
Myers \& Swiatecky 1996), however BPAL12 EOS is still able to sustain the 
measured mass 1.44 $M_\odot$ of the pulsar PSR1916+13
as the maximum gravitational mass of non--rotating neutron stars constructed
with this EOS is $1.46$~\msun. 

(2) Stiff: {\it Sahu et al. (1993)}:
This is a field theoretical EOS for neutron matter in beta equilibrium based 
on the chiral sigma model.  The model includes an isoscalar vector field 
generated dynamically and reproduces the empirical values of the nuclear matter
saturation density and binding energy and also the isospin symmetry
coefficient for asymmetric nuclear matter.  The energy per nucleon of nuclear
matter according to Sahu et al. (1993) is in very good agreement, up
to about four times the equilibrium nuclear matter density, with estimates 
inferred from heavy--ion collision experimental data.  The maximum 
gravitational mass of non--rotating neutron stars constructed with this 
EOS is $2.59$~\msun.

For our computations, we constructed the composite EOS for the 
entire span of neutron star densities by joining the
new high density EOS models to that of Negele \& Vautherin (1973) for the
density range ($10^{14} - 5\times10^{10}$) \gcc, Baym et al. (1972) for 
densities down to $\sim 10^3$ \gcc and Feynman et al. (1949) for densities 
less than $10^3$ \gcc.

\section{Results and Discussion}

The equilibrium sequences of rotating neutron stars depend on two parameters: 
the central density ($\epsilon_c$) and the rotation 
rate ($\Omega$).  For purpose of illustration, we choose three limits
in this parameter space. These are: (i) the static or non--rotating limit, 
(ii) the limit at which instability to quasi--radial mode sets in and 
(iii) the centrifugal mass shed limit. The last limit corresponds to the 
maximum $\Omega$ for which centrifugal forces are able to balance the inward 
gravitational force.

Table 2 summarizes the non--rotating neutron star structure parametes for the 
EOS models BBB1, BBB2, BPAL21 and BPAL32. The values listed correspond to the 
maximum stable mass configuration.  The entries in this table 
are the central density ($\epsilon_c$), the gravitational mass ($M_G$), 
the rest (baryonic) mass ($M_0$) of the neutron star 
and the radius ($R$). The maximum mass is an indicator of the softness/stiffness
of the EOS and its values as listed in Table 2 reflect that the EOS models used
are all intermediate in stiffness.  Among these, BPAL21 is the softest EOS  
and BPAL32 the stiffest one.

In Table 3 we list the following quantities corresponding to the maximum 
gravitational mass 
configurations: central density, rotation rate, moment of 
inertia ($I$), gravitational mass, ratio of rotational kinetic energy  to 
total graviational energy ($T/W$), equatorial radius, eccentricity ($e$) where 
$e$ is defined to be $\sqrt{1 - R_p^2/R^2}$, $R_p$ being the polar radius of the
configuration, ratio 
of rotation rate ($\omega_c$) of stellar fluid relative to the inertial frame 
at the centre of the star to the rotation rate, the angular momentum ($J$), 
the value of the radius  of the innermost stable orbit ($r_{orb}$),  the polar,
 forward  and backward  redshifts ($Z_p$, $Z_f$, $Z_b$), the rest mass ($M_0$) 
 and the proper mass ($M_p$) of the neutron star.
The values of these quantities are listed in Table 4 for the maximum angular 
momentum models. From Table 3, it can be seen that the gravitational mass of 
the maximum stable rotating configuration has a value that is close to 
$2$ \msun. Interestingly, this is close to predictions from analysis of LMXB 
observational data (Zhang et al. 1996). Therefore, if the internal constitution 
of the compact star in LMXBs were to be described by the EOS models that we 
have considered here, and furthermore, if it were to have a mass 
$\sim 2$ \msun, the star has to be rotating near centrifugal break--up speeds
(rotation periods $\sim 0.5$ ms). Interestingly, for such configurations, it 
can be seen from Table 3, that the separation of the innermost stable orbit 
from the neutron star surface (namely, the boundary layer extent) is highly 
EOS dependent. This separation  ranges 
from $0.07$ km  to $1.031$ km. This relatively large spread can be understood 
in terms of the spreads on values of $\epsilon_c$ and $\Omega$ corresponding 
to the centrifugal mass shed. These results will have relevance in modeling 
LMXBs/QPOs.  For such neutron stars as the central objects in  LMXBs, there 
will be a very significant re--ordering of the contributions of the disk and 
the boundary layer luminosities to the total luminosity (Thampan \& Datta 1997).

The results of our computations for rotating neutron stars corresponding to
the four new EOS  models are given below. 

\subsection {EOS Model BPAL21}

The normal sequences for this EOS have rest mass 
$M_0$/\msun $ < 1.9395$ and supramassive sequences have rest mass 
$1.9395 < M_0$/\msun $ < 2.2515$.  

\subsection {EOS Model BBB1}

In Fig. 3 we show the functional dependence of the gravitational mass with
central density.  In this and all subsequent figures, the bold solid curve
represents the non--rotating or static limit, and the bold dashed curve the 
centrifugal mass shed limit.  The long dashed curve is the constant--$\Omega$ 
sequence corresponding to the period $P=1.558$~ms.  The thin solid curves that
are roughly horizontal are the constant rest mass evolutionary sequences. 
The evolutionary sequences above the maximum stable non-rotating mass
configuration are the supramassive evolutionary sequences, and those that are 
below this limit are the normal evolutionary sequences.  The almost vertical
thin dashed line is the limit for instability against quasi--radial modes. The
supramassive evolutionary sequences beyond the quasi--radial mode instability
limit are represented by dotted lines. 
The maximum mass sequence for this EOS  corresponds to a rest mass value of
$2.471$ \msun.  The supramassive sequences lie in the rest mass range of
$2.356$ \msun$ < M_0 < 2.471$ \msun.  The gravitational mass of the maximum 
stable rotating  configuration is  $2.135$ \msun and its radius is $13.129$~km.
If we assume that the fastest pulsar known to date, 
PSR 1937+21, has the canonical mass value of 1.4 \msun and is described by EOS 
model BPAL21, then this neutron star should have a central density of about 
$1.2\times10^{15}$ \gcc.  

In Fig. 4  we give a plot of $M_G$ as a function of $R$.  For the millisecond
pulsar PSR 1937 + 21 with an assumed  mass value of $1.4$ \msun, this 
corresponds to  a radius  of  $11$ km.

In Fig. 5 we display the plot of $\Omega$ as a function of the specific angular 
momentum $cJ/GM_{0}^{2}$.  The inset shows a close--up view of the region 
surrounding  the instability limit to quasi--radial mode near the centrifugal 
mass shed limit.  It is clear from this figure that the maximum mass 
rotating 
model (represented by the plus sign) has a lower angular velocity than the 
maximum--$\Omega$ model (represented by the intersection of the line 
representing the instability to quasi--radial modes with that of the 
centrifugal mass shed limit).  

\subsection {EOS Model BBB2}

The equilibrim sequences for BBB2 are displayed on Figs. 6 -- 8 for the same 
representative set of parameters as for EOS model BBB1

For this EOS, the supramassive sequences have rest masses between $2.261$ \msun
and $2.653$ \msun and the maximum mass at mass shed limit is $2.272$ \msun with
an equatorial radius of $12.519$~km.

\subsection {EOS Model BPAL32}

This EOS model being the stiffest out of the four models that we consider 
here, has the highest value for the maximum rotating gravitational mass 
($2.3$ \msun).  The rest masses of supramassive sequences lie 
in the range $2.263$ \msun $< M_0 < 2.655$ \msun.

\noindent Since the behaviour of $M$ with $\epsilon_c$ and $R$ and
that of $\Omega$ with $cJ/GM_0^2$ for EOS models BPAL21 and BPAL32 are more 
or less similar to those for EOS models BBB1 and BBB2, we do not display the 
corresponding figures for the former EOS models here.

In Table 5 we list the values of the various parameters for the constant
$\Omega$ sequences for the four EOS models considered in this work.
In general, $r_{orb}$ exhibits three characteristics: (a) $r_{orb}$ is 
non--existent (b) $r_{orb} < R$, and (c) $r_{orb} > R$.  For the first
two cases, $r_{orb}$ is taken to be the Keplerian orbit radius at the
surface of the star.  From Table 5 it can be seen that for low central 
densities, stable orbits can exist all the way up to the surface of the
neutron star but for high enough central densities, the boundary layer (the 
separation between the surface of the neutron star and its innermost stable 
orbit) can be substantial ($\sim 5$ km for the maximum
value of the listed central densities). These results will have applications
in modeling accretion flows in LMXBs.

We now make a brief reference to other similar work.  For equilibrium 
Keplerian angular velocity corresponding to the period of millisecond 
pulsar PSR 1937+21 and an assumed mass of $1.4$ \msun for the neutron
star, Friedman et al. (1986) suggest that stiff EOS for neutron
star interior are favoured.  A similar conclusion but based on a pulsar 
glitch model and the crustal moment of inertia considerations  has  
been reported by Datta \& Alpar (1993).  The work of Friedman et al. (1986) 
show that for a given EOS, the models with maximum gravitational mass also 
have the greatest frequency of rotation.  Cook et al. (1994) found that while 
models with maximum gravitational mass also (due to stability conditions defined 
by Friedman et al. 1988) have the maximum rotation rate $\Omega$,
the models for maximum gravitational mass and maximum-$\Omega$ do not in
general coincide.  In particular, for EOS models that display causality
violation near or before the maximum stable mass non--rotating configuration,
the maximum-$\Omega$ model occurs before (in central density and $\Omega$)
the maximum mass model at the mass shed limit.
The EOS models that we have considered  here do not violate the causality 
condition until well beyond the maximum stable mass non--rotating 
configuration.  Our computations show that the maximum gravitational mass 
rotating models for these EOS occur (in central density and $\Omega$) before 
the maximum--$\Omega$ models. In view of absence of correlation between QPO
frequency and source count rate (as suggested by  recent observations, see
Berger et al. 1996; Zhang et al 1996), our results on marginally stable
Keplerian orbits corresponding to realistic EOS will have application in
understanding kilo--Hertz quasi--periodic oscillations in X--ray binaries 
in terms of strong--field general relativity, where rapid rotation of the 
accreting neutron star is important.



\begin{thebibliography}{}

\bibitem{} Backer D. C., Kulkarni S. R., Heiles C., Davis M. M., Goss W. M.,  
1982, Nat 300, 615

\bibitem{} Baldo M., Bombaci I., Burgio G. F., 1997, A\&A 328, 274, (BBB) 

\bibitem{} Bardeen J. M., 1970, ApJ 162, 71

\bibitem{} Bardeen J. M., 1972, ApJ 178, 347

\bibitem{} Bardeen J. M., Wagoner R. V., 1971, ApJ 167, 359

\bibitem{} Baym G., Pethick C. J., Sutherland P. G., 1972, ApJ 170, 299

\bibitem{} Berger M., van der Klis M., van Paradijs J., et al., 1986, ApJ 469, 
           L13

\bibitem{} Blaizot J. P., 1980, Phys. Rep.  64, 171.

\bibitem{} Bombaci I., 1995, Perspectives on Theoretical Nuclear Physics. 
           In: Bombaci I., Bonarccorso A., Fabrocini A., et al. (eds.) Proc.
           VI Convegno su Problemi di Fisica Nucleare Teorica, ETS, Pisa, 
           p.223

\bibitem{} Bombaci I., 1996, A\&A 305, 871 

\bibitem{} Bombaci I., Lombardo U., 1991, Phys.Rev. C44, 1892 

\bibitem{} Bonazzola S., Schneider J., 1974, ApJ 191, 273

\bibitem{} Bonazzola S., Gougoulhon E., Salgado M., Marck J. A., 1993, A\&A 278, 421

\bibitem{} Brown G. E., Bethe H. A., 1994, ApJ 432, 659

\bibitem{} Butterworth E. M., 1976, ApJ 204 561

\bibitem{} Butterworth E. M., Ipser J. R., 1976, ApJ 204, 200

\bibitem{} Carlson J., Pandharipande V. R., Wiringa R. B., 1983, Nucl. Phys. 
A401, 59

\bibitem{} Cook G. B., Shapiro S. L.,  Teukolsky S. A., 1994, ApJ 424, 823

\bibitem{} Datta B., Alpar M. A.,  1993, A\&A 275, 210

\bibitem{} Datta B., Thampan, A. V., Bombaci, I.,  1997, in preparation

\bibitem{} Friedman J. L., Ipser J. R.,  Parker L., 1986, ApJ 304, 115

\bibitem{} Friedman J. L., Ipser J. R.,  Sorkin R. D., 1988, ApJ 325, 722

\bibitem{} Feynman R. P., Metropolis N.,  Teller E., 1949, Phys. Rev. 
75, 1561

\bibitem{} Hachisu I., 1986, ApJS 61, 479

%
\bibitem{} Komatsu H., Eriguchi Y.,  Hachisu I., 1989, MNRAS 237, 355

\bibitem{} Lacombe M., Loiseau B., Richard J. M., et al., 1980, Phys. Rev. C21,
            861
 
\bibitem{} Lattimer J., Pethick C., Prakash M., Haensel P., 1991, 
           Phys. Rev. Lett. 66, 2701

\bibitem{} Machleidt R., 1989, Adv. Nucl. Phys. 19, 189

\bibitem{} Myers W. D., Swiatecky W. J., 1996, Nucl. Phys. A601, 141.

\bibitem{} Negele J. W.,  Vautherin D., 1973, Nucl. Phys. A207, 298

\bibitem{} Page D., Applegate J.H.,  1992, ApJ 394, L17

%
\bibitem{} Sahu P. K., Basu R.,  Datta B., 1993, ApJ 416, 267

\bibitem{} Salgado M., Bonazzola S., Gourgoulhon E., Haensel P., 1994a, 
           A\&A 291, 155

\bibitem{} ------------- 1994b, A\&AS 108, 455

\bibitem{} Schiavilla R., Pandharipande V. R., Wiringa R. B., 1986, Nucl. Phys.
 A449, 219 

\bibitem{} Skyrme T. H. R., 1956, Philos. Mag. 1, 1043 
%

\bibitem{} Taylor J. H., Weisberg J. M., 1989, ApJ 345, 434

\bibitem{} Thampan A. V., Datta  B., 1998,  MNRAS (in press)

\bibitem{} Wiringa R. B., Smith R. A., Ainsworth T. L., 1984, Phys. Rev. C29, 
           1207

\bibitem{} Wiringa R. B., Fiks V., Fabrocini A., 1988, Phys. Rev. C38, 1010

\bibitem{} Zhang W., Lapidus I., White N. E., Titarchuk L., 1996, ApJ 469, L17

\end{thebibliography}
\end{document}